\documentclass{PoS}
\usepackage{subcaption}

\title{Results from the DM-Ice17 Dark Matter Experiment at the South Pole}

\ShortTitle{Results from the DM-Ice17 Dark Matter Experiment at the South Pole}

\author{\speaker{Jay Hyun Jo}, on behalf of the DM-Ice Collaboration\\
        Department of Physics and Wright Laboratory, Yale University\\
        E-mail: \email{jayhyun.jo@yale.edu}}

\abstract{DM-Ice is a phased experimental program using low-background NaI(Tl) crystals with the aim to unambiguously test the claim of dark matter detection by the DAMA experiments. DM-Ice17, consisting of 17 kg of NaI(Tl), has been continuously operating at a depth of 2457 m in the South Pole ice for over five years, demonstrating the feasibility of a low-background experiment in the Antarctic ice. Studies of low and high energy spectra, an annual modulation analysis, and a WIMP exclusion limit based on the physics run of DM-Ice17 are presented. We also discuss the plan and projected sensitivity of a new joint physics run, COSINE-100, with upgraded detectors at the Yangyang Underground Laboratory in Korea.}

\FullConference{38th International Conference on High Energy Physics\\
		3-10 August 2016\\
		Chicago, USA}

\begin{document}

\section{Introduction}
Astrophysical and cosmological observations provide strong evidence that the dark matter constitutes nearly a quarter of the Universe~\cite{DM1, DM2}. The weakly interacting massive particle (WIMP) is a theoretically favored to explain this dark matter~\cite{WIMP}. One method of detecting WIMP is to measure the annual modulation of a WIMP signal caused by the Earth's motion in galactic rest frame, with period of one year. Only the DAMA/NaI and DAMA/LIBRA experiments, located at Laboratori Nazionali del Gran Sasso in Italy, claim to have observed an annual modulation of dark matter accurate to 9.3~$\sigma$~\cite{DAMA_Combined, DAMA_LIBRA}, which is in conflicts with several experiments~\cite{LUX}.

The DM-Ice experiment aims to resolve this tension by operating an experiment in the Southern Hemisphere using the same target material, thallium-doped sodium iodide (NaI(Tl)) scintillating crystals. While the expected dark matter modulation has the same phase everywhere on Earth, any other modulating environmental backgrounds will have 180$^\circ$ out of phase between the Northern and Southern Hemisphere. By operating at the South Pole, we expect that DM-Ice can disentangle the dark matter phase from seasonal variations. 

\section{DM-Ice17 Detector}

As the first stage of the DM-Ice experimental program, DM-Ice17 was built to demonstrate the feasibility of performing low background measurements in the Antarctic ice~\cite{DMIceConcept}. DM-Ice17 consists of two 8.47~kg NaI(Tl) detectors, referred to as Det-1 and Det-2, which were deployed 2450~m into the South Pole ice in December 2010~\cite{DMIceFirst}. The physics run started on June 2011 and ended on January 2015 with a total exposure of 60.8~kg$\cdot$yr. The detectors have continued stable operations since.

\section{Results of DM-Ice17}

\begin{figure}
	\centering
	\begin{subfigure}[b]{0.49\textwidth}
		\centering
		\includegraphics[width=\textwidth]{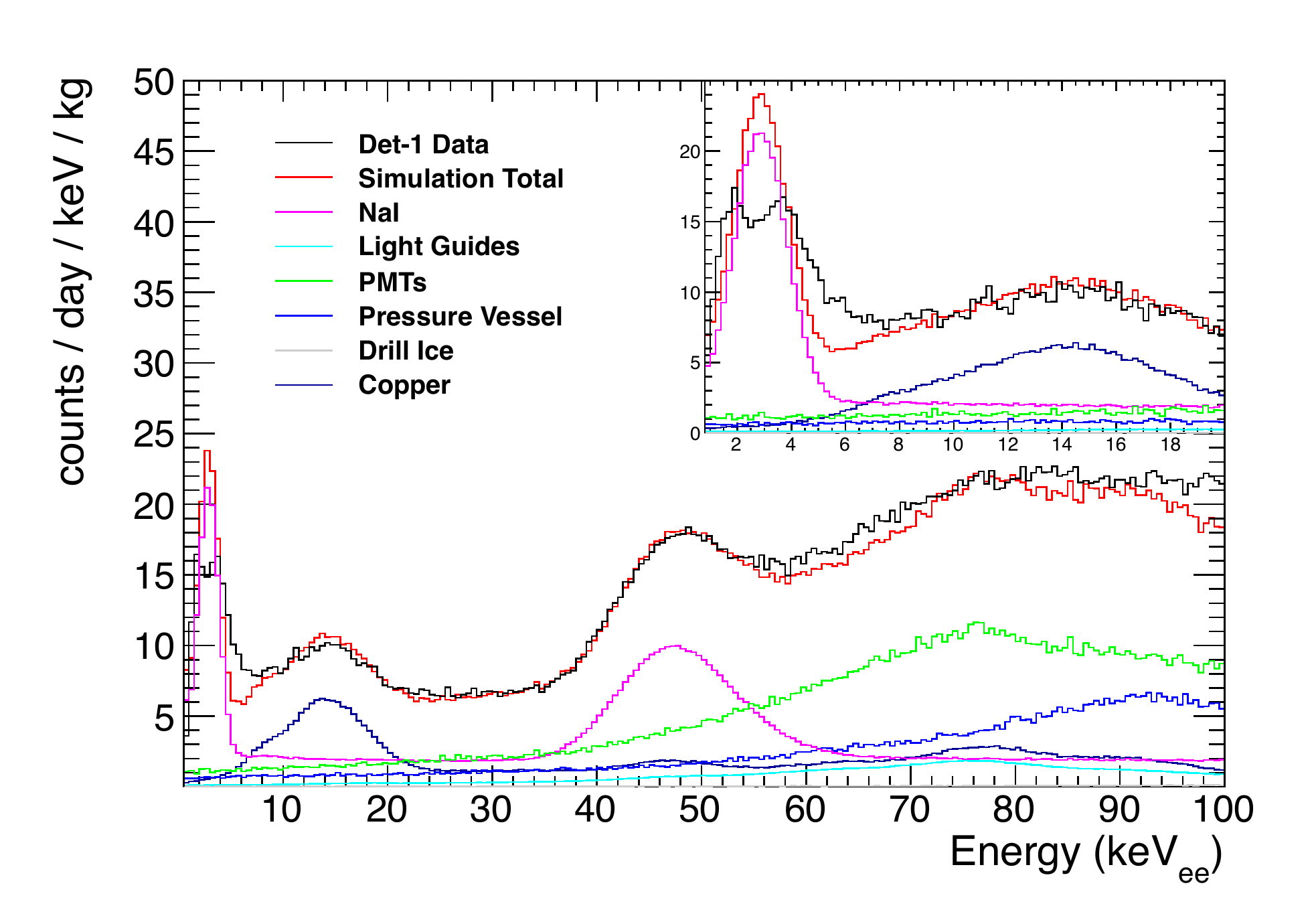}
	\end{subfigure} 
	\hfill
	\begin{subfigure}[b]{0.49\textwidth}
		\centering
		\includegraphics[width=\textwidth]{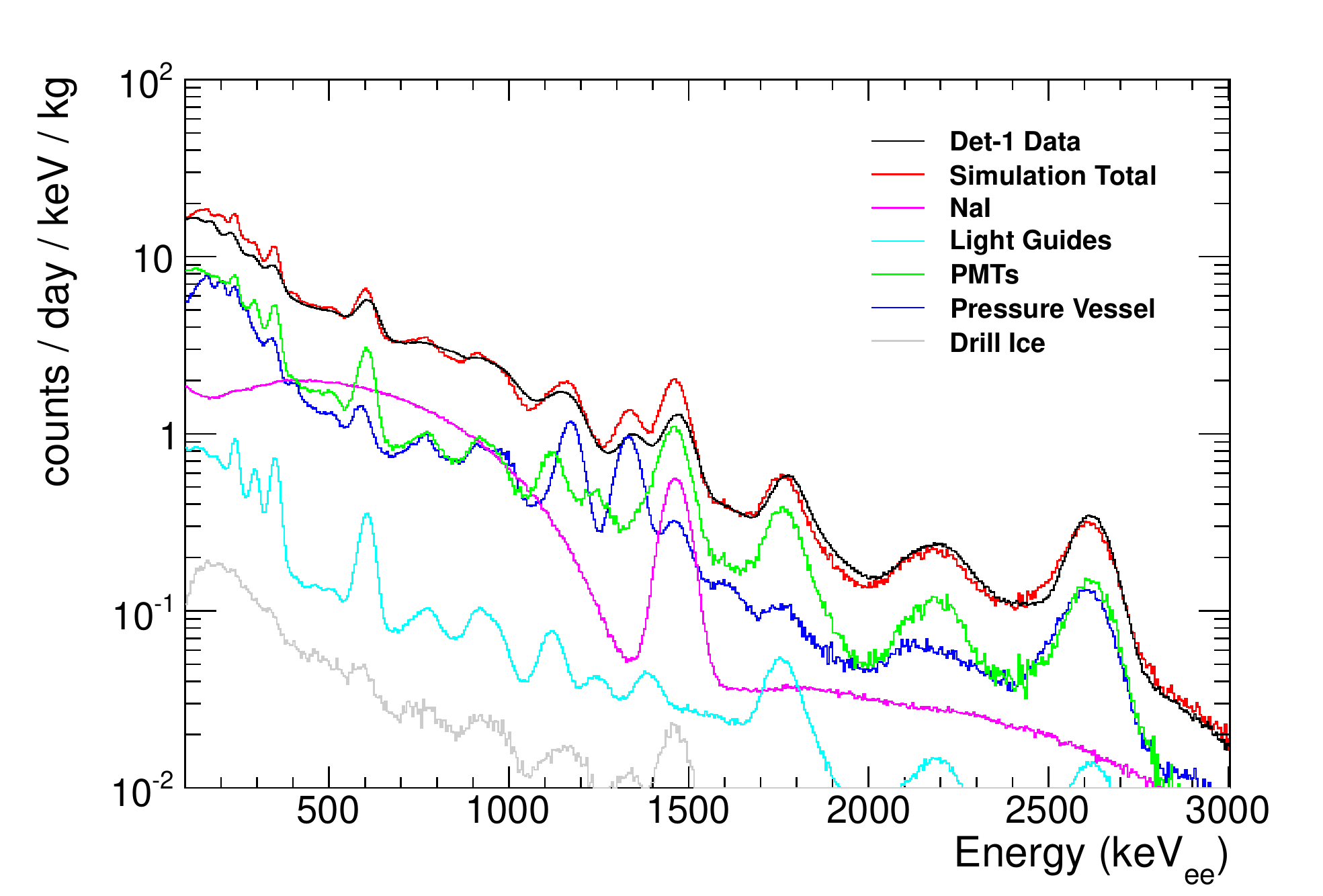}
	\end{subfigure}
	\hfill
	\caption{DM-Ice17 Det-1 background data at low (left) and high (right) energies. Detector components with main contaminations are simulated. Background peaks are used for energy calibrations.}
	\label{fig:background}
\end{figure}

DM-Ice17 is calibrated with intrinsic and cosmogenic backgrounds, such as $^{210}$Pb and $^{125}$I, respectively. Calibration measurements were performed separately within the low and high energy regions, due to non-linear NaI light responses. Despite inability to use external calibration sources, the $^{125}$I cosmogenic peak confirms the calibration by identifying both its energy and the decay time, with the expected half-life of 59.4 days. 

Simulated background model produced with Geant4 are consistent with the data from DM-Ice17, as shown in Fig.~\ref{fig:background}~\cite{DMIceFirst}. At the extreme end of the low energy region, below 4~keV, simulation does not agree with the data, mainly due to the efficiency of signal retention during noise removal. Thus, an analysis threshold for Det-1 and Det-2 was set to 4~keV and 6~keV, respectively.

As confirmed by simulations, the dominant sources of contamination are from the $^{40}$K, $^{238}$U and $^{232}$Th chains in the crystals, PMTs, and pressure vessels. The 3~keV peak is due to Auger electrons and x-rays from $^{40}$K decays in the crystals. A broad peak at 14~keV can be attributed mostly to surface contamination of the $^{238}$U-chain in the copper encapsulation, which has been observed previously in other NaI(Tl) experiments~\cite{ANAIS}. The flat background of the crystal reaching to 30 keV is dominated by contributions from $^{210}$Pb and $^{40}$K.

\begin{figure}
	\centering
	\begin{subfigure}[b]{0.48\textwidth}
		\centering
		\includegraphics[width=\textwidth]{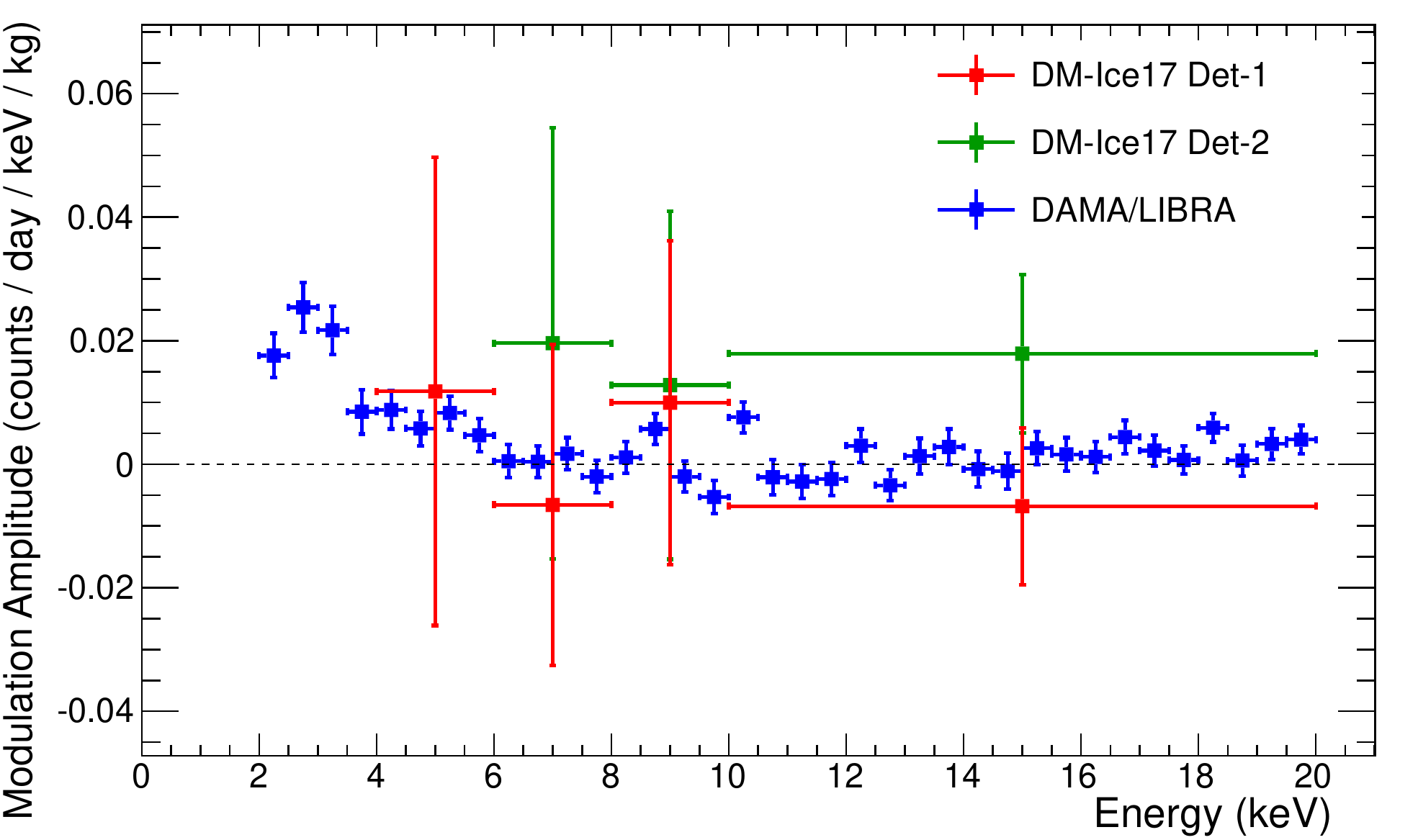}
	\end{subfigure}
	\hfill
	\begin{subfigure}[b]{0.51\textwidth}
		\centering
		\includegraphics[width=\textwidth]{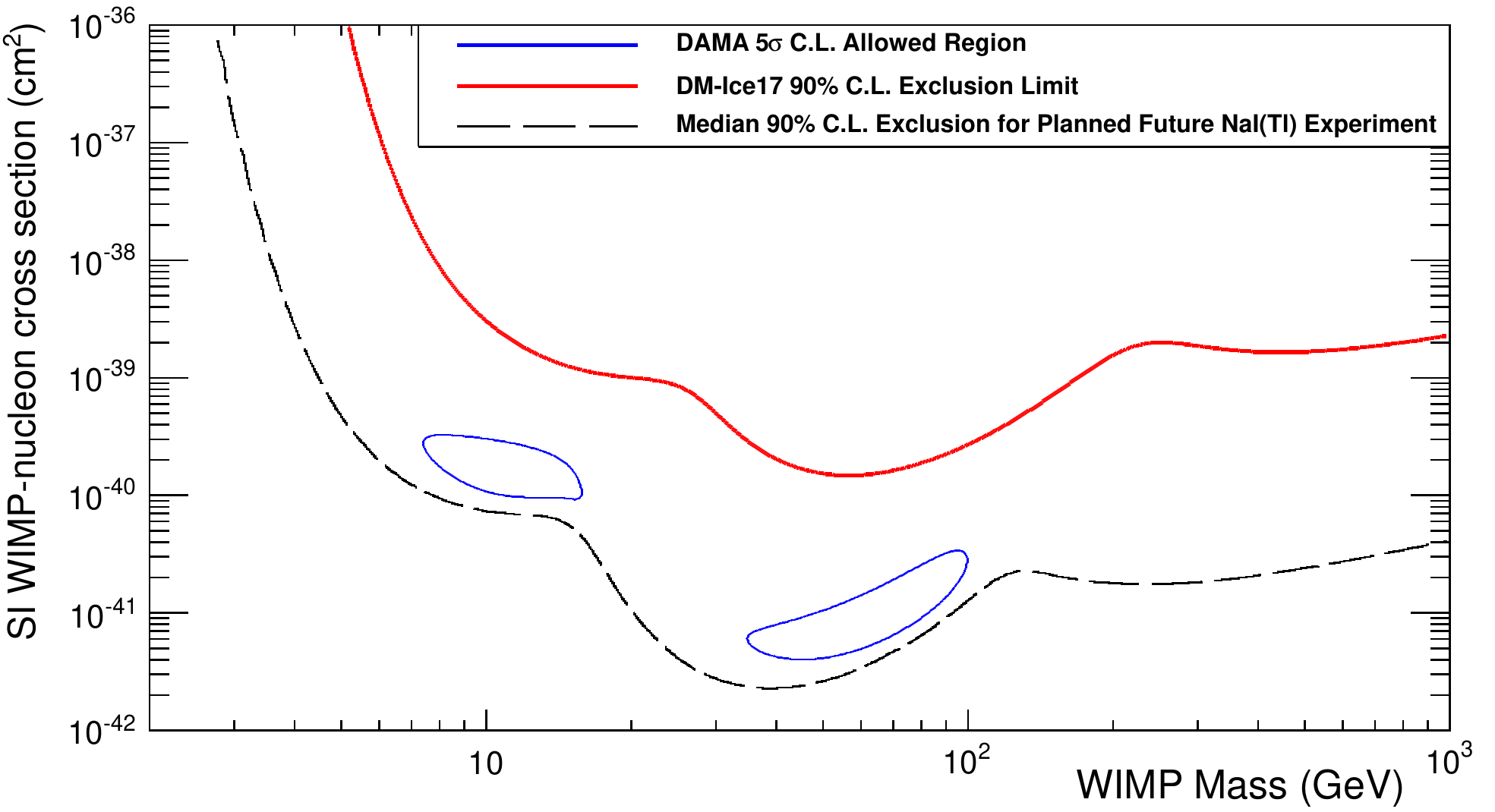}
	\end{subfigure}
	\caption{Left: Det-1 (red) and Det-2 (green) modulation amplitudes of each energy bin compared to the values from DAMA/LIBRA. Right: WIMP exclusion limit at 90\% C.L. from the 60.8~kg$\cdot$yr DM-Ice17 physics dataset (red), with DAMA preferred 5$\sigma$ C.L. contour (blue) for comparison.}	
	\label{fig:modulationResult}
\end{figure}

Maximum likelihood fits of the background subtracted event rates for 4 different energy bins (4--6, 6--8, 8--10, and 10--20~keV) have been performed within the DM-Ice17 annual modulation analysis. When modulation amplitudes of each energy bin are compared to the values from DAMA/LIBRA with period and phase fixed to that of an expected dark matter signal (1 year and 152.5 days, respectively), as shown in Fig.~\ref{fig:modulationResult} (left), it reveals that the data from DM-Ice17 are consistent with both the null hypothesis and DAMA/LIBRA signal under the limitations of the detector~\cite{DMIceAnnMod}. However, the result provides the strongest limit in the Southern Hemisphere by a direct detection dark matter search (See Fig.~\ref{fig:modulationResult} (right)).

\section{Prospect: COSINE-100}

\begin{figure}
	\centering
	\begin{subfigure}[b]{0.49\textwidth}
		\centering
		\includegraphics[width=0.9\textwidth]{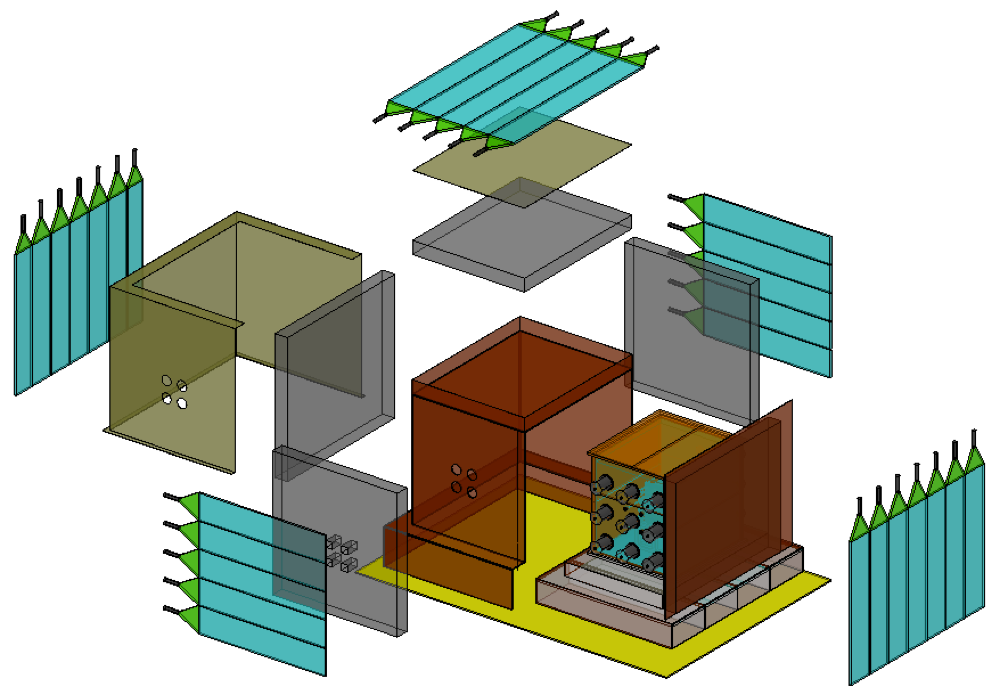}
	\end{subfigure}
	\begin{subfigure}[b]{0.49\textwidth}
		\centering
		\includegraphics[width=\textwidth]{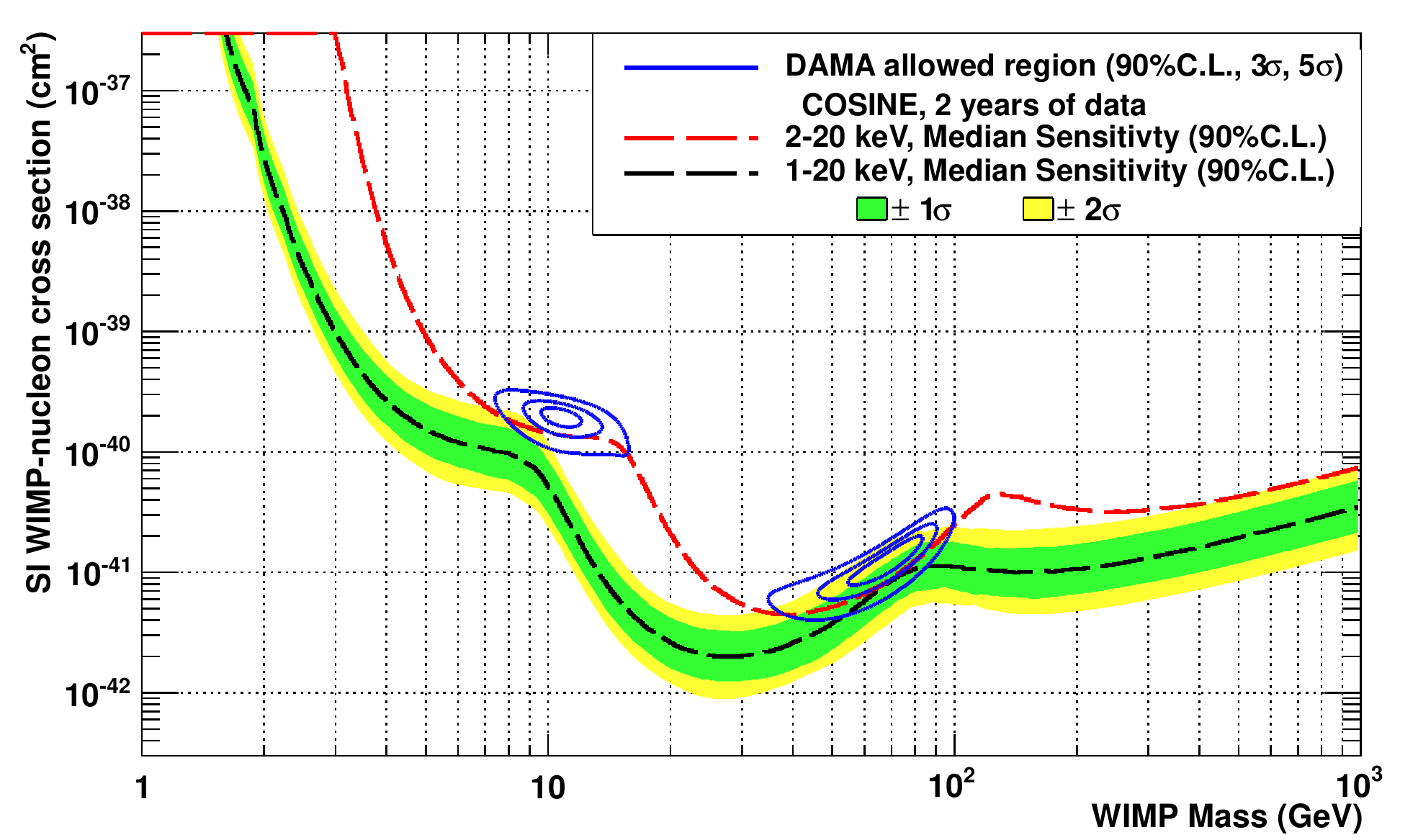}
	\end{subfigure}

	\caption{Left: COSINE-100 shielding structure. Right: Projected sensitivity for COSINE-100 with 1~keV (2~keV) threshold in black (red) with 2 years of data.}
	\label{fig:cosine}
\end{figure}

DM-Ice17 is limited by small exposure and high backgrounds, and targeted R\&D programs have been ongoing to overcome this limitation~\cite{Walter, KIMS_NaI}. Cleaner crystals with more mass can be accessible with a new crystal vendor\footnote{http://www.alphaspectra.com} and newer PMTs with lower background and higher quantum efficiency are needed.

DM-Ice and KIMS-NaI have formed a new collaboration, COSINE-100, located at Yangyang underground laboratory in South Korea. COSINE-100 consists of 8 NaI(Tl) crystals with a total mass of 106~kg and a 2000 liter liquid scintillator veto, to help reduce low energy backgrounds by tagging $^{40}$K events. Figure~\ref{fig:cosine} (left) shows the shielding structure of COSINE-100 which includes 3~cm of copper, 20~cm of lead, and 3~cm of 37 plastic scintillator panels for cosmic ray muon tagging. Data taking for COSINE-100 began in September 2016. 

Figure~\ref{fig:cosine} (right) shows projected sensitivity of COSINE-100, assuming 2--4 counts/day/keV/kg flat background, depending on crystal powder type. It is expected that within two years of running, COSINE-100 will achieve a sensitivity to test DAMA's result.

\section{Conclusion}

DM-Ice17, the only dark matter detector in the Southern Hemisphere, is operating successfully under Antarctic ice and established the South Pole as a site for underground low-background experiments. The DM-Ice17 physics run data, taken over 3.6 years for a total exposure of 60.8~kg$\cdot$yr, shows no evidence of an annual modulation in the 4--20~keV energy range. Results give the strongest exclusion limit in the Southern Hemisphere, but are consistent with both the null hypothesis and DAMA's results as DM-Ice17 is limited by exposure time and intrinsic background rates.

COSINE-100 will utilize more massive NaI(Tl) crystals with lower intrinsic backgrounds with the aid of a liquid scintillator veto. COSINE-100 has been taking data since September 2016 and two years of data will be able to test DAMA's assertion for the detection of dark matter.

\begin{acknowledgments}
The author would like to thank the conference organizers for the invitation at ICHEP2016. The DM-Ice collaboration thanks the Wisconsin IceCube Particle Astrophysics Center (WIPAC) and the IceCube collaboration for their on-going experimental support and data management. This work was supported in part by NSF Grants No.~PLR-1046816, PHY-1151795, and PHY-1457995, WIPAC, the Wisconsin Alumni Research Foundation, and Yale University. 

\end{acknowledgments}

\end{document}